\documentclass[12pt]{article}
\usepackage{amsmath,amssymb,graphicx,mathrsfs,hyperref,slashed}

\newcommand{\be}{\begin{equation}}
\newcommand{\ee}{\end{equation}}
\newcommand{\bea}{\begin{eqnarray}}
\newcommand{\eea}{\end{eqnarray}}

\newcommand{\wh}{\widehat}
\newcommand{\wt}{\widetilde}

\def\({\left(} \def\){\right)}
\def\N{{\cal N}}

\begin{document}

\title{\vspace{-1.8in}
\vspace{0.3cm} {Restoring predictability \\ in semiclassical gravitational collapse}}
\author{\large Ram Brustein${}^{(1)}$,  A.J.M. Medved${}^{(2)}$ \\
 \hspace{-1.5in} \vbox{
 \begin{flushleft}
  $^{\textrm{\normalsize
(1)\ Department of Physics, Ben-Gurion University,
    Beer-Sheva 84105, Israel}}$
$^{\textrm{\normalsize (2)  Department of Physics \& Electronics, Rhodes University,
  Grahamstown 6140, South Africa }}$
 \\ \small \hspace{1.7in}
    ramyb@bgu.ac.il,\  j.medved@ru.ac.za
\end{flushleft}
}}
\date{}
\maketitle

 \begin{abstract}

Hawking showed that the radiation emitted from a classical collapsing shell is thermal. Here, we show that a semiclassical collapsing shell emits radiation that is only approximately thermal, with small but significant deviations. The most important difference is the presence of small off-diagonal elements in the radiation density matrix with a magnitude of order $1/\sqrt{S_{BH}}$, $S_{BH}$ being the Bekenstein-Hawking entropy  of the incipient black hole. The off-diagonal elements store the correlations between the collapsing shell and the emitted radiation and  allow information to continuously leak from the collapsed body. The rate of escape of information is initially very small, increasing to order one by the Page time, when the black hole  has lost half its original entropy. We show that,  until the Page time, the radiation is almost exactly thermal and that, from this time on, it begins to purify at an increasing rate.

\end{abstract}
\newpage

\section{Introduction}

Hawking's remarkable calculation  showing that a  black hole (BH) emits thermal radiation \cite{Hawk}  leads to the information-loss paradox \cite{info}. (See also, {\it e.g.}, \cite{info2,info3,info4}.) This puzzle  has attracted a  recent surge of interest thanks, in large part, to a closely related conundrum  that was
first clarified  in \cite{AMPS} and is commonly known as the firewall paradox. (See \cite{Sunny1,Braun} for earlier, related ideas  and  \cite{fw1,fw2,fw3,fw4,Sunny2,avery,lowe,vv,pap} for a sample of the ensuing debate.)

Hawking's seminal result \cite{info} is clear: A BH, even if formed out of a pure state of matter,  emits radiation as if it were a perfect black body. The  radiation arrives at infinity in a thermal state of maximum entropy and so the radiation can contain no information about the original matter. Later on, Hawking's result and the information-loss paradox or the ``BH information problem" have been rephrased in information-theoretic terms.

It is sometimes said  that the information loss is not problematic until such time as the BH has evaporated away. This is, however, not quite accurate. There will inevitably come a time when the BH reaches a critically small mass where the Hawking calculation, which presumes a large BH in a slowly evolving spacetime, must break down. When this occurs, there will be a ``remnant'' of the BH that no longer has the size to account for the missing information. And, even if the idea of Planck-sized hyperentropic  objects is still floated out there as a viable resolution, the general consensus is that such exotic states of matter would undermine the stability of the vacuum or come with another exorbitant price tag. (See, {\it e.g.}, \cite{remnant}.)

Regardless of the status of the paradox, it has become widely accepted that the idea that information could somehow be lost cannot be correct. One argument in this direction that we find particularly convincing is the following: Let us consider a finite-energy and  finite-size BH, which could be as large as the largest astrophysical BH's. If this BH is to be considered as a physical state that can be formed from known matter and can decay to known matter, then it  must be described by a quantum state in some Hilbert space, just as any other matter system would be. If this argument is accepted, then classical, geometric notions like the horizon being a rigid surface of infinite red-shift and the singularity being a point of infinite curvature can no longer be valid. We have described some supporting evidence to this effect in \cite{flucyou}, where it was shown that quantum fluctuations of the background geometry tame  classical infinities that are associated with the horizon. Another piece of supporting evidence was provided in \cite{RJ}, where it was shown that the horizon of a finite BH is subject to fluctuations due to the uncertainty principle  and, as such, will have a degree of transparency that is missed in a perfectly classical treatment \cite{RM,RB}. For earlier papers that specifically consider horizon fluctuations, see \cite{fluc1,fluc2,fluc3,fluc4,fluc5}.

The importance of treating BH's as quantum states has been recently advocated by Dvali and Gomez \cite{Dvali1} (with Lust \cite{Dvali2}), as well as the current authors with collaborators \cite{RM,RJ,RB,flucyou}. Let us emphasize, however, that what we mean by ``quantum" in this context is  only the leading semiclassical correction. We argue that it is not necessary to invoke Planck-scale physics to describe some aspects of the physics of macroscopic BH's; in particular, those aspects that are related to Hawking's result.

Once it is accepted that a radiating BH should be viewed as a quantum state, then it should also be accepted that the burning of a BH should be governed by the same principles that govern the burning of a neutron star or the  burning of an encyclopedia. So that the burning of a BH cannot involve new layers of unpredictability. But what is left to discover is the correct mechanism for the release of information. This has turned out to be a highly non-trivial task, as should be obvious given the  decades of debate.
Indeed, it can be inferred from the information-theory treatment of Page \cite{page} that a  perturbative treatment of the matter cannot do the job.

Page charted the progress of an initially pure state representing the BH, which he then divides into two separate parts: One part represents  the  slowly radiating BH and the other, the emitted radiation. He assumed that the initial pure state is a completely random state and  showed that the first bit of information only becomes available to an external observer when the  BH  has lost half of its original entropy to the emitted radiation.  Until this so-called ``Page time'', most of the information is stored in either correlations within the evaporating body or else in correlations between the body and the (early) radiation. After the Page time, the rate of release of information quickly becomes of order unity. So that, by the end of the evaporation process, all the information does get released, just as it would for any other unitary process.

Later, Hayden and Preskill \cite{HaydenPreskill} complemented the Page analysis by a similar discussion. Also assuming an completely random initial state,
they tracked the fate of the information content of a matter state that is thrown into an existing BH and found similar results to those of Page.
If the  matter is thrown into a ``young" BH, its information content
takes about a Page time to get released and, if thrown into an ``old" BH, the
information about the matter  is released almost instantaneously.

The challenge is then to look for an information-release mechanism that is compatible with the previous framework and that  allows for the total escape of information before the BH has had the chance to completely evaporate. In \cite{RB}, it was proposed that the origin of the BH information paradox is the use of a  strictly classical geometry for the BH.  It was further proposed that the leading semiclassical corrections to classical values should be taken into account by assigning a quantum wavefunction to the BH; thus incorporating the semiclassical  fluctuations of the background geometry. In other words, one should calculate expectation values by using  this wavefunction rather than the classical geometry. Here, we implement the program of \cite{RB} and propose a concrete  mechanism that is indeed based on the semiclassical fluctuations of the background geometry.

In spite of using the small effect of a fluctuating background geometry, our treatment of matter is intrinsically non-perturbative in the following way: Had we asked how  equally small fluctuations  of the matter fields impact upon Hawking's calculation, the method would be  doomed from the start. This is because
the redshift that leads to the thermal spectrum is already an exponentially large quantity, meaning that small matter  corrections will be exponentially suppressed. This aspect was discussed by Hawking \cite{Hawk} and allowed him to argue forcefully that his result is robust against  such quantum fluctuations. What we rather do is reveal a previously overlooked consequence of a fluctuating semiclassical background.

The semiclassical fluctuations of the geometry lead to a small correction to  the form of the density matrix for the emitted radiation. The corrected matrix agrees with Hawking's on the diagonal but includes off-diagonal contributions as well. As will be shown, this structure in the density matrix  allow information to continuously leak from the BH.  The rate of escape of information is initially very small, suppressed by $1/S_{BH}$ and continuously increases, reaching order one by the Page time. This $1/S_{BH}$  suppression is quite strong for a macroscopic BH; however, it is much weaker than the exponential suppression $e^{-S_{BH}}$ that was found by Page. The reason for the higher rate in our calculation is that we use a specific quantum state for the collapsing shell, while  keeping track of the correlations between the emitted radiation and the collapsing shell. Page, on the other hand, did not consider the correlations and tracked the information content of just the radiation. We then show that, up  until the Page time, the radiation is almost exactly thermal in spite of the correlations and that, from this time onward, it begins to purify at an increasing rate, consistent with unitary evaporation.

\subsection{Conventions}

Our units are chosen such that Planck's constant $\hbar$ and Newton's constant $G$ are made explicit, whereas  all other fundamental constants are set to unity.

We assume a four-dimensional Schwarzschild BH  (generalizations are
straightforward) of
large but finite mass $M\gg\sqrt{\hbar/G}$, with the usual metric
$\; ds^2 = -\left(1-\frac{R_{S}}{r}\right)dt^2+ \left(1-\frac{R_{S}}{r}\right)^{-1}dr^2
+d\Omega^2_2\;$.
Here, $\;R_{S}=2MG\;$  denotes the position of the horizon radius.
We also  use the Schwarzschild ``tortoise'' coordinate
$\; r^{\ast}=\int^r\;dr \sqrt{-g^{tt}g_{rr}} = r+R_{S}\ln(r-R_{S})\;$
and the associated pair of null coordinates,
$\;v =t+r^{\ast}\;$ and  $\;u=t-r^{\ast}\;$.

For a Schwarzschild  BH, the values of its Hawking temperature and Bekenstein--Hawking
entropy are respectively $\;T_{H}=\frac{\hbar}{4\pi R_{S}}\;$, $\;S_{BH}=\frac{\pi R_{S}^2}{\hbar G}\;$.

A dot indicates a derivative with respect to $t$.

\section{Mechanism for information transfer}

Page's analysis \cite{page} or, equivalently, the assumption of unitarity,  makes it clear that information must be transferred from the BH to the emitted Hawking radiation. But what has  been missing is a mechanism for this transfer in the context of Hawking's original calculation \cite{Hawk,info}. We will eventually  show  how semiclassical quantum fluctuations of the collapsing shell can serve this purpose, but first let us briefly recall Hawking's basic framework (also see \cite{Ford}).

\subsection{Radiation emitted by a classical collapsing shell}

To begin, one considers a collapsing shell of matter and focuses on a certain class of null rays; those that enter the interior of the shell when  the shell's (classical) radius  $R_{shell}$ is significantly larger than its Schwarzschild radius $R_{S}$ but  exit the interior at a time close to that  of horizon formation $\;(R_{shell}-R_{S})_{in}\gg (R_{shell}-R_{S})_{out}\;$. Because the ray exits when $\ (R_{shell}-R_{S})_{out}/R_S\ll 1\;$, it  undergoes a large redshift after passing through the interior region. From the perspective of an asymptotic observer, this strong time-dependence leads to  particle creation.  Hawking's calculation made this description quantitative, allowing him to use the radiation emitted from a collapsing shell as a model of particle emission from a slowly evaporating BH that was formed by collapsing matter.

As an initial step, one describes the ray's outgoing null coordinate $u$ in terms of the shell trajectory $R_{shell}=R_{shell}(T)$ (where $T$ is Minkowski time for the interior region) and then uses matching conditions at the surface and the center of  the shell  to relate this expression to the advanced time difference $v_0-v$. Here,  $v$ is the ingoing null coordinate for the ray and $v_0$ is the (finite) value of $v$  at the moment of  horizon formation.

This matching procedure and the use of geometric optics enabled Hawking to deduce the form of the outgoing asymptotic modes ({\it i.e.}, those that arrive at future null infinity ${\cal I}^{+})$ when traced back to past null infinity ${\cal I}^{-}$. These ``out-modes'' can then be related to the ``in-modes" originating at ${\cal I}^{-}$ through a Bogolubov transformation. After this, a Fourier transformation of the out-modes determines  the form of the Bogolubov coefficients $\alpha_{\omega^{\prime}\omega}$, $\beta_{\omega^{\prime}\omega}$ as  functions of the ingoing and outgoing mode frequencies (respectively, $\omega^{\prime}$ and $\omega$) integrated over $v$. The remainder of the calculation entails  using analytic continuation and  then applying the Bogolubov normalization
condition. At the end, one has isolated the ``negative-energy'' Bogolubov coefficients $\beta_{\omega^{\prime}\omega}$, which serve as a measure of particle creation as seen
by an asymptotic observer.

The Bogolubov transformation between in- and out-modes
goes as (with angular-momentum quantum numbers suppressed)
\be
F_{\omega} \;=\; \int\limits_{0}^{\infty} \;d\omega^{\prime}\;\Big(\alpha_{\omega^{\prime}\omega}f_{\omega^{\prime}}
+\beta_{\omega^{\prime}\omega}f^{\ast}_{\omega^{\prime}}\Big)\;,
\label{prestuff}
\ee
where $F_{\omega}$
is a ``traced-back'' out-mode and
$\;f_{\omega^{\prime}}=\frac{1}{\sqrt{2\pi}}e^{i\omega^{\prime}v}\;$
is a basis function for an in-mode.

The Hawking (single-particle) density matrix for the out-modes can then  be
expressed as
\bea
\rho_H(\omega,\wt{\omega})
&=&\int\limits_{-\infty}^{v_0} dv \int\limits_{0}^{\infty} d\omega^{\prime} \int\limits_{0}^{\infty} d\omega^{\prime\prime}
\;\beta^{\ast}_{\omega^{\prime}\omega}\beta_{\omega^{\prime\prime}\widetilde{\omega}}
 \frac{e^{iv\left(\omega^{\prime}-\omega^{\prime\prime}\right)}}{2\pi}\;,
\label{rhoH}
\eea
where the left-hand side is a short-hand notation for
$\langle 0_{in}|\rho_H(\omega,\wt{\omega})|0_{in}\rangle$. The state of matter $|0_{in}\rangle$ denotes the initial vacuum: the state that is annihilated by positive-frequency in-modes.

The Bogolubov coefficients are related by  $\;\beta_{\omega^{\prime}\omega}\sim - i\alpha^{\ast}_{(-\omega^{\prime})\omega}\;$ and are evaluated by an ingenious method of ray tracing, relying on the geometric optics of the modes as they traverse across the shell  \cite{Hawk,Ford}. The result is
\bea
\;\beta_{\omega^{\prime}\omega}&\propto& \frac{1}{2\pi}\int\limits_{-\infty}^{v_0}dv\ e^{i \omega' v}\ e^{-i 2 \omega  R_S \ln(v_0-v)}
\cr &=& \Gamma\left(1-i2\omega R_{S}\right) \left(i\omega^{\prime}\right)^{-1+i2\omega R_{S}}
\frac{1}{2\pi} e^{iv_0\left(\omega^{\prime}-\omega\right)}\;.
\label{betaH}
\eea
The logarithm in the first line of Eq.~(\ref{betaH}) takes into account the discontinuity in the phase of the modes as they pass across the shell at an advanced time $v$ close to $v_0$, when the shell  is near its Schwarzschild radius.

The full explicit expression for the integrand in the Hawking density matrix~(\ref{rhoH}) is the following:
\bea
\beta^*_{\omega'\omega}\beta_{\omega''\wt{\omega}}
&=&\frac{t^*_{\omega} t_{\wt{\omega}}}{(2\pi)^2} \frac{1}{(\omega\wt{\omega})^{1/2}}
\Gamma\left(1+i\frac{\omega}{\kappa}\right) \Gamma\left(1-i\frac{\wt{\omega}}{\kappa}\right)
(\omega')^{-1/2-i\frac{\omega}{\kappa}}
(\omega'')^{-1/2+i\frac{\wt{\omega}}{\kappa}}\hspace{.3in} \cr && \times\ e^{-\frac{\pi}{\kappa}\frac{\omega+\wt{\omega}}{2}} \ e^{iv_0(\omega''-\wt{\omega}-\omega'+\omega)}\;.
\label{eqbeta}
\eea
Here, $t_{\omega}$ is the transmission coefficient through the gravitational potential barrier (see \cite{info}) and we have introduced the surface curvature $\;\kappa= 2\pi T_H/\hbar = 1/ (2 R_S)\;$.

The expression in Eq.~(\ref{eqbeta}) does not have the correct dimensions when $\;\omega\ne\wt{\omega}\;$. The dimensionality of $\rho_H$
should be $\omega^{-1}$, so that its trace
$\;{\rm Tr}\;\rho_H(\omega,\wt{\omega}) = \int d\omega \; \rho_H(\omega,\omega)\;$  is dimensionless.  The actual dimensionality is off from the required power
of  $-1$ by a power of $-i(\omega-\wt{\omega})/\kappa$. We can formally fix this by multiplying the left-hand side by
$\kappa^{i(\omega-\wt{\omega})/\kappa}$, which will be implicitly assumed.

To verify  that the matrix is diagonal, one can start by noticing the fact that the basis functions $\frac{1}{\sqrt{2\pi}}e^{i\omega v}$ are orthogonal on the interval $-\infty<v<v_0$,
$\;\int\limits_{-\infty}^{v_0} dv\; \frac{e^{iv\left(\omega^{\prime}-\omega^{\prime\prime}\right)}}{2\pi}
=\delta(\omega^{\prime}-\omega^{\prime\prime})\;$.
The density matrix then includes the relevant factor
\bea
\rho_H(\omega,\wt{\omega}) &\sim &
\int\limits_{0}^{\infty} d\omega^{\prime} \; (\omega^{\prime})^{-1+i4\pi R_{S}(\widetilde{\omega}-\omega)}\; \nonumber \\
& =& \int\limits_{-\infty}^{\infty} dy \; e^{iy\left[4\pi R_{S}(\widetilde{\omega}-\omega)\right]}
\nonumber \\
&=& \frac{1}{2R_{S}}\delta(\widetilde{\omega}-\omega)\;.
\label{freqint}
\eea

The three integrations over $w$, $w'$ and $v$ have then conspired
to act like  an exact delta function on the outgoing frequencies. The final result is simply a delta function times a thermal spectrum,
\bea
\rho_H(\omega, \widetilde{\omega})
 &\propto& \frac{1}{e^{\frac{\hbar\omega}{T_{H}}}-1} \delta(\omega-\widetilde{\omega})\;.
\label{spec}
 \eea

Equation~(\ref{spec}) exposes the thermal nature of the radiation, but what is really central to the information-loss problem is the absence  of off-diagonal terms in the density matrix, as this is where the correlations between different modes could be stored.

\subsection{Wavefunction of a semiclassical collapsing shell}

In  Subsection~2.3, we will recall  how the Bogolubov coefficients are evaluated from the discontinuity of the modes across the shell and then  determine how this discontinuity is modified for the case of a  semiclassical shell. This will require knowing the    wavefunction of the collapsing shell of matter at times when it is about to enter its Schwarzschild radius. The wavefunction will be the focus of the current subsection, and we begin here by returning our attention to the shell's radial trajectory.

At times $T$ close to (but preceding) the time of  horizon formation $T_0$,
this trajectory  takes the form \cite{Ford}
\be
R_{shell}(T) \;=\; R_{S} + A\left(T_0-T\right) +{\cal O}\left[(T_0-T)^2\right]\;,
\label{shell}
\ee
where $A$ is a dimensional constant.
The aforementioned matching conditions enables one to translate this relation into  Schwarzschild coordinates for the outgoing null ray,
\be
t\;=\;-R_{S}\ln\left[T_0 -T\right]+\cdots\;,
\label{tout}
\ee
\be
r^{\ast}\;=\;+R_{S}\ln\left[R_{shell}-R_{S}\right]+\cdots\;,
\label{rout}
\ee
where the ellipses indicate constants as well as corrections to these near-horizon forms.

From an operational perspective,  one is  constructing a reference system out of $T$ and $R_{shell}$ coordinates and using them  to describe the  trajectory of an  out-mode. Observers taking measurements at different times are expected to  agree on the path followed. This would be the case in a perfectly classical spacetime.

But the quantum situation is  different. There will now be an inherent quantum fuzziness associated with the shell trajectories resulting from the quantum fluctuations of the shell.  The standard protocol of tracing over the shell and working with the reduced density matrix for the matter is supposed to trivialize this concern. However,  as first proposed in \cite{RB} and discussed at length in \cite{flucyou}, this procedure can be expected to have non-trivial consequences on semiclassical physics, particularly when probing near a BH horizon.

As  the geometry of interest is determined by the collapsing shell of matter, we need to assign  the shell  a quantum wavefunction $\Psi_{shell}$. Since considerations are for a spherical collapse, we will assign the shell a wavefunction that only depends on its radius, $\;\Psi_{shell}=\Psi_{shell}(R_{shell})\;$. In general, the wavefunction of the shell will be time dependent. For example, its average position should decrease with time towards its Schwarzschild radius. However, we are interested in the shell when its size is near its Schwarzschild radius and  (as before)  in high-frequency modes that are not sensitive to the relatively slow rate at which the shell evolves  in time. It follows that we can approximate by using a static wavefunction for the shell.

What form of static wavefunction should be assigned to the shell? Clearly the average radial position of the shell has to be its classical radius. As the classical radial position of the shell approaches its Schwarzschild radius  $\;R_{shell}\to R_{S}\;$, it should obey $\;\langle \Psi_{shell}| R_{shell} |\Psi_{shell} \rangle \to R_S\;$.   The leading semiclassical extension enters through the width of the wavefunction: The quantum spread about the average value. We have found, as will become clear later on, that to evaluate the leading correction to the Hawking density matrix it suffices to to specify the width of wavefunction. Hence, it  is not necessary to specify the wavefunction beyond the leading semiclassical order. We will therefore assign the shell the simplest wavefunction that can be described by its average and width; a Gaussian wavefunction with average $R_S$ and width $\sigma^2$,
\be
\left.\Psi_{shell}(R_{shell})\right|_{R_{shell}\to R_{S}}\; = \;{\cal N}^{-1/2} e^{-\frac{\left({R}_{shell}-R_{S}\right)^2}{2\sigma^2}}
\label{prewave}
\;.
\ee

The value of $\sigma^2$ can be determined as follows. On dimensional grounds, we expect that the width of the shell can be as small as the Planck scale  but not less, $\;\sigma^2 \sim \hbar G =  l_p^2\;$. (Here, $l_p$ is the Planck length.) We will adopt a width that is proportional to this minimal value for reasons that will be explained shortly. The exact proportionality constant will not be very important for our conclusions about the semiclassical correction to the Hawking density matrix. However, in the following we do propose a way to determine the constant.

Adopting the the line of reasoning presented in \cite{RM,RB} (also see \cite{flucyou}), we can determine the proportionality constant such that $\sigma^2 = \hbar G/\pi $. Let us see how this conclusion is reached.
The shell is itself almost a BH at the relevant times, and so we expect that the shell's  wavefunction  goes smoothly to that of the BH, $\;\Psi_{shell} \to \Psi_{BH}\;$, as the shell approaches its Schwarzschild radius $\;R_{shell}\to R_{S}\;$. Here, $\Psi_{BH}$ means  the semiclassical wavefunction for an already formed BH.
As discussed in \cite{RB} , the wavefunction of the spherically symmetric mode of a semiclassical BH can be  described by the Gaussian wavefunction
$\;\Psi_{BH}(R)\;=\;{\cal N}^{-1/2}e^{-\frac{({R}-R_{S})^2}{2C_{BH} R_{S}^2}}\;$.
Here,   $\;{\cal N} = \int_0^{\infty}d{R}\;4\pi{R}^2
\; e^{-\frac{({R}-R_{S})^2}{C_{BH} R_{S}^2}}\;$
ensures that the wavefunction's probability
is normalized to unity and the parameter $C_{BH}$ is discussed directly below. For future reference,
$\;{\cal N}=4\pi R_{S}^2\sqrt{\pi C_{BH} R_{S}^2}\;$ up to highly suppressed corrections.

The width of the wavefunction can then be  determined by the general arguments of \cite{RM} relying on the application of Bohr's correspondence principle. As also explained in \cite{RB,flucyou}, the parameter that determines the width of the BH wavefunction is
\be
C_{BH}\; = \; \frac{\hbar G}{\pi R_{S}^2}=\frac{1}{S_{BH}}\;.
\label{lambda}
\ee
This parameter is a small but finite number for a semiclassical BH, the inverse of the Bekenstein--Hawking entropy. It was explained in \cite{RB,flucyou} that  the
parameter $C_{BH}$ is the natural measure of classicality in a BH spacetime. For physical quantities, semiclassical deviations from a strictly classical spacetime  typically start at order $C_{BH}$. Hence, a physical consequence of a wavefunction whose quantum spread is $C_{BH}$ can be expected to appear at this order.

In light of the above discussions,  we arrive at the final form that will
be  used  as the shell wavefunction,
\be
\left.\Psi_{shell}(R_{shell})\right|_{R_{shell}\to R_{S}}\; = \;{\cal N}^{-1/2} e^{-\frac{\left({R}_{shell}-R_{S}\right)^2}{2C_{BH} R_{S}^2}}
\label{wave}
\;.
\ee

To evaluate the semiclassical density matrix,
we need to translate the probability distribution in terms of $R_{shell}$ to a probability distribution in terms of the advanced time $v$.  Taking  the limit $\;R_{shell}\to R_S\;$ (with the understanding that $R_{shell}$ fluctuates quantum mechanically)
and comparing the  $R_{shell}(T),T$ coordinates to the $u,v$ coordinates, we know that
$\;\frac{v_0-v}{R_S} \propto e^{- u/( 2R_S)}\;$.
Then, since $\;u=t-r^{\ast}\;$ and $\;t\sim -r^{\ast}\;$ near the horizon ({\em cf}, Eqs.~(\ref{tout},\ref{rout})), it follows that $\;\frac{v_0-v}{R_S} \sim e^{r^{\ast}/ R_S }\;$. But, also near the horizon, $\;r^{\ast}\sim R_S \ln\left(\frac{r-R_S}{R_S}\right)\;$ and so $\;\frac{v_0-v}{R_S} \sim e^{\ln\left(\frac{r-R_S}{R_S}\right)}\;$.  It
can now be concluded that, for $\;R_{shell} \sim R_S\;$,
\be
R_{shell}-R_S \;\simeq\; v_0-v\;
\ee
and
\be
R_{shell}^2\;\simeq\; R_S^2 +2R_S(v_0-v)+(v_0-v)^2\;.
\ee

It follows that expectation values of operators $\wh{O}(R_{shell})$ can be evaluated using a convolution with the probability distribution according to the standard rules of quantum mechanics,
\be
\hspace{-2in}\langle  \wh{O}(R_{shell})\rangle\;=\;
\frac{4\pi}{\N} \int\limits_0^\infty dR_{shell}\; R_{shell}^2\ e^{-\frac{(R_{shell}-R_S)^2}{2 \sigma^2}} {O}(R_{shell})
\nonumber
\ee
\be\hspace{0.4in} \;\simeq\; \frac{4\pi}{\N}  \int\limits_{-\infty}^{\infty} dv_{shell}\left[ R_S^2 + 2R_S(v_0-v_{shell})+(v_0-v_{shell})^2\right]\ e^{-\frac{(v_0-v_{shell})^2}{2 \sigma^2}}{O}(v_{shell})\;,
\label{vfunction}
\ee
where $\;\sigma^2=R_S^2C_{BH}/2\;$ and
the limits of the $v_{shell}$ integral are estimated as follows: Since $\;v_{shell}\sim R_S-R_{Shell}\;$ and $R_{shell}$ goes from $0$ to $\infty$,  the range of $v_{shell}$ is $\;-\infty <v_{shell}< R_S\;$. However, because $\;R_S^2\gg \sigma^2\;$, it makes almost no difference if we set the upper limit to $\infty$.

\subsection{Semiclassical corrections to Hawking's density matrix}

The goal is now to understand how $\Psi_{shell}$ impacts upon the Hawking calculation.  The main reason that the  wavefunction for the shell changes the density matrix is that the phase discontinuity of the  out-mode across the shell changes.  We continue, as assumed previously, to use the in vacuum $|0_{in}\rangle$ as the state of matter. The total state of the matter and the shell is assumed to be in a product state
\be
|\Psi_{shell,matter}\rangle\;=\;|\Psi_{shell}\rangle |0_{in}\rangle \;.
\ee

From  Eq.~(\ref{betaH}), we see that the discontinuity in the phase is $\ln(v_0-v)$. However, in the quantum case, the position of the shell is at $v_{shell}$ rather than at $v_0$. Thus the only relevant quantities that are sensitive to the shell wavefunction are the Bogolubov coefficients $\beta$. It follows that the single-particle density matrix that corrects the Hawking density matrix is given by
\be
\rho_{SC}(\omega,\wt{\omega})
\;=\;\int\limits_{-\infty}^{v_0} dv \int\limits_{0}^{\infty} d\omega^{\prime} \int\limits_{0}^{\infty} d\omega^{\prime\prime}
\langle\Psi_{shell}|\;\beta^{\ast}_{\omega^{\prime}\omega,~SC}~ \beta_{\omega^{\prime\prime}\widetilde{\omega},~SC}|\Psi_{shell}\rangle \frac{e^{iv\left(\omega^{\prime}-\omega^{\prime\prime}\right)}}{2\pi}\;,
\label{rhoSCD}
\ee

To evaluate the ``semiclassical'' coefficients $\beta_{\omega^{\prime\prime}\widetilde{\omega},SC}$, we use the same method of ray-tracing as used to evaluate the Hawking density matrix. As just explained, the only difference is that the discontinuity in the phase is  at $v_{shell}$ rather than at $v_0$. Hence, the semiclassical $\beta$'s will change according to
\bea
\;\beta_{\omega^{\prime}\omega,~SC}&\propto& \frac{1}{2\pi} \int\limits_{-\infty}^{v_{shell}}dv\ e^{i \omega' v}\ e^{-i 2 \omega  R_S \ln(v_{shell}-v)}
\cr &=& \Gamma\left(1-i2\omega R_{S}\right) \left(i\omega^{\prime}\right)^{-1+i2 \omega R_{S}}
\frac{1}{2\pi} e^{iv_{shell}\left(\omega^{\prime}-\omega\right)}+\cdots\;,
\label{betaQ}
\eea
where the ellipsis stands for terms that lead to higher-order corrections in $C_{BH}$.

When compared to the original classical coordinates, the phase discontinuity across the shell is now slightly different; and so the phase of $\beta_{SC}$ is  slightly different than in the classical case.

The resulting semiclassical correction to the density matrix can be determined as follows: One needs  to substitute  expression (\ref{betaQ}) for $\beta_{SC}$  into Eq.~(\ref{rhoH}) for the density matrix. Then the main source of change appears in the scalar product of the two basis functions for the in-modes.

 Let us recall that, in the classical case, this scalar product yields $\;I_C=\frac{1}{2\pi}\int\limits_{-\infty}^{v_0} dv \; e^{i v (\omega'-\omega'')}=\delta(\omega'-\omega'')\;$. Whereas, semiclassically, this outcome should  be revised to
\be
{I}_{SC}(\omega'-\omega'';v_{shell})=\frac{1}{2\pi}\int\limits_{-\infty}^{v_0} dv\;  e^{i (v-v_{shell}) (\omega'-\omega'')}\;.
\ee
One  can then changes variables to $\;v^\prime=v-v_{shell}\;$ and find that
\be
{I}_{SC}(\omega'-\omega'';v_{shell})=\frac{1}{2\pi}\int\limits_{-\infty}^{v_0-v_{shell}} dv^\prime \;
e^{i v^\prime (\omega'-\omega'')}\;.
\ee

The next step is to evaluate the expectation value of $\wh{I}_{SC}$ by convolving $I_{SC}(\omega'-\omega'';v_{shell})$ with the probability distribution for the shell's wavefunction according to the rule in Eq.~(\ref{vfunction}),
\bea
\langle \wh{I}_{SC}(\omega'-\omega''; v_{shell}) \rangle & =&
\frac{4\pi}{\N}\int\limits_{-\infty}^{\infty} d v_{shell} \left[ R_S^2 + 2R_S (v_0-v_{shell}) +(v_0-v_{shell})^2\right] \nonumber  \\
&\;&\;\;\;\;\;\times\; e^{-\frac{(v_0-v_{shell}^2)}{2 \sigma^2}}
  \frac{1}{2\pi}\int\limits_{-\infty}^{v_0-v_{shell}} dv^\prime \; e^{i v^\prime (\omega'-
\omega'')}\;.
\label{gaussbdry}
\eea

We can, however,  split $\langle \wh{I}_{SC}(\omega'-\omega''; v_{shell}) \rangle$ into its classical
(delta-function)  and  semiclassical contributions,
\bea
\langle \wh{I}_{SC}(\omega'-\omega''; v_{shell}) \rangle&= & \left\langle\frac{1}{2\pi}\int\limits_{-\infty}^{0} dv^{\prime}\;  e^{i v^{\prime} (\omega'-\omega'')}+\frac{1}{2\pi}\int
\limits_{0}^{\wt{v}} dv^{\prime}\;  e^{i v^{\prime} (\omega'-\omega'')}\right\rangle \cr &=& \delta(\omega'-\omega'')+ \left\langle\frac{1}{2\pi}\int\limits_{0}^{\wt{v}} dv^{\prime}\;  e^{i v^{\prime} (\omega'-\omega'')}\right\rangle,
\eea
where $\wt{v}=v_0-v_{shell}$.
The delta-function term will simply  lead to Hawking's density matrix, meaning that  the semiclassical correction to the Hawking result is given by
\be
\Delta I_{SC}(\omega'-\omega'')=\frac{4\pi}{\N}\int\limits_{-\infty}^{\infty} d \wt{v} \left[ R_S^2 + 2R_S\wt{v}+\wt{v}^2\right] e^{-\frac{\wt{v}^2}{2 \sigma^2}} \frac{1}{2\pi} \int\limits_{0}^{\wt{v}} dv^{\prime}\;  e^{i v^{\prime} (\omega'-\omega'')}\;.
\label{j2}
\ee

First performing the right-most integral,
\bea
\Delta I_{SC}(\omega'-\omega'')&=&\frac{4\pi}{\N}\int\limits_{-\infty}^{\infty} d \wt{v} \left[ R_S^2 + 2R_S\wt{v}+\wt{v}^2\right] e^{-\frac{\wt{v}^2}{2 \sigma^2}} \frac{1}{2\pi} \left[\frac{e^{i\wt{v}(\omega'-\omega'')}-1}{i(\omega'-\omega'')}\right]
\cr &=&
\frac{4\pi}{\N}\int\limits_{-\infty}^{\infty} d \wt{v} \left[ R_S^2 + 2R_S\wt{v}+\wt{v}^2\right] e^{-\frac{\wt{v}^2}{2 \sigma^2}}
\frac{1}{2\pi} e^{i\frac{\wt{v}}{2}(\omega'-\omega'')} \frac{\sin\left[ \frac{\wt{v}}{2}(\omega'-\omega'')\right]}{\left(\frac{\omega'-\omega''}{2}\right)}\;, \ \hspace{.5in}
\label{j3}
\eea
and then changing variables to $\;\wt{V}=\wt{v}-i \frac{\omega'-\omega''}{2}\sigma^2\;$, we have
\bea
\Delta I_{SC}(\omega'-\omega'')&=&\frac{4\pi}{\N}\int\limits_{-\infty}^{\infty} d \wt{V} \left[ R_S^2 + 2R_S\left(\wt{V}+i \frac{\omega'-\omega''}{2}\sigma^2\right)+\left(\wt{V}+i \frac{\omega'-\omega''}{2}\sigma^2\right)^2\right]
 \cr &\times&
e^{-\frac{\wt{V}^2}{2 \sigma^2}}\ e^{-\frac{1}{8}(\omega'-\omega'')^2 \sigma^2}\ \frac{1}{2\pi}
\frac{\sin\left[ \left(\wt{V}+i \frac{\omega'-\omega''}{2}\sigma^2\right) \frac{(\omega'-\omega'')}{2}\right]}{\left(\frac{\omega'-\omega''}{2}\right)}\;.
\label{j4}
\eea

Now, using the fact that the integration domain of $\wt{V}$ and the Gaussian in $\wt{V}$ are both even, we need only to pair even terms in the left-most square bracket with even terms in the expansion of the $\sin$ function (see below) and odd terms  with odd, respectively. In addition, we  only  require net contributions that  are even in $\omega'-\omega''$
 because odd ones  would vanish when the frequencies are later integrated.

Next consider that, to leading order in the semiclassical expansion (zeroth and first order in $\;\sigma^2\sim C_{BH}\;$),
\bea
&& R_S^2 + 2R_S\left(\wt{V}+i \frac{\omega'-\omega''}{2}\sigma^2\right)+\left(\wt{V}+i \frac{\omega'-\omega''}{2}\sigma^2\right)^2 \;=\; \cr && R_S^2+2 R_S\wt{V} + i R_S \sigma^2 (\omega'-\omega'')+\cdots \;,
\eea
which needs to be combined with
\bea
&& \sin\left[ \left(\wt{V}+i \frac{\omega'-\omega''}{2}\sigma^2\right)\frac{\omega'-\omega''}{2}\right] \;=\;  \cr && \sin\left(\wt{V}\frac{\omega'-\omega''}{2}\right) \cos\left(i \frac{(\omega'-\omega'')^2}{4}\sigma^2\right) +\cos\left(\wt{V}\frac{\omega'-\omega''}{2}\right)
\sin\left(i \frac{(\omega'-\omega'')^2}{4}\sigma^2\right)\;. \ \hspace{.5in}
\eea
Then, taking into account the additional factor $\frac{1}{\pi(\omega'-\omega'')}$ and the requirements that the integrand be even in both $\wt{V}$ and  $\omega'-\omega''$, we find only two relevant terms,
\bea
&& +2R_S \wt{V}\sin\left(\wt{V}\frac{\omega'-\omega''}{2}\right) \cosh\left( \frac{(\omega'-\omega'')^2}{4}\sigma^2\right) \cr &&
- R_S \sigma^2 (\omega'-\omega'')\cos\left(\wt{V}\frac{\omega'-\omega''}{2}\right)
\sinh\left( \frac{(\omega'-\omega'')^2}{4}\sigma^2\right).
\label{sinhcosh}
\eea

 The second of the previous terms yields the integral
\be
R_S \sigma^2 (\omega'-\omega'') \frac{4\pi}{\N}\int_{-\infty}^{\infty} d \wt{V}\ e^{-\frac{\wt{V}^2}{2 \sigma^2}}\
\cos\left(\wt{V}\frac{\omega'-\omega''}{2}\right)\;=\;\frac{1}{R_S} \sigma^2 (\omega'-\omega'') \ e^{-\frac{(\omega'-\omega'')^2}{8} \sigma^2}\
\ee
times the accompanying sinh function, whereas the first term yields  the same
integral,
\be
 2R_S \frac{4\pi}{\N}\int_{-\infty}^{\infty} d \wt{V}\
e^{-\frac{\wt{V}^2}{2 \sigma^2}}\
\wt{V}\sin\left(\wt{V}\frac{\omega'-\omega''}{2}\right)\;=\;\frac{1}{R_S} \sigma^2 (\omega'-\omega'') \ e^{-\frac{(\omega'-\omega'')^2}{8} \sigma^2}\;,
\ee
times the cosh function. But, because of the sign difference in Eq.~(\ref{sinhcosh}),
this leads to the subtraction of the $\sinh$ function  from the $\cosh$,
leaving only the decaying exponential terms in the sum.

So all together, to leading order,
\bea
&&\Delta I_{SC}(\omega'-\omega'')\;=\;\frac{1}{R_S} \sigma^2 (\omega'-\omega'') \frac{1}{\pi(\omega'-\omega'')}e^{-\frac{(\omega'-\omega'')^2}{4}\sigma^2} \cr &&\times  \left[ \cosh\left( \frac{(\omega'-\omega'')^2}{4}\sigma^2\right) -\sinh\left( \frac{(\omega'-\omega'')^2}{4}\sigma^2\right) \right] \cr &=&
\frac{1}{\pi R_S}  \sigma^2 e^{-\frac{(\omega'-\omega'')^2}{2}\sigma^2}\;.
\label{j5}
\eea
 Then, after substituting $\;\sigma^2=\frac{1}{2} C_{BH} R_S^2\;$, we obtain
\bea
\Delta I_{sc}(\omega'-\omega'')&=& \frac{1}{2\pi} R_S C_{BH}\ e^{-\frac{(\omega'-\omega'')^2}{4} R_S^2 C_{BH}}\;.
\label{j6}
\eea

The semiclassical correction to the density matrix is then given by the explicit expression,~\footnote{We have used coordinate freedom to
eliminate the remaining dependence on $v_0$.}
\bea
&&\rho_{SC}(\omega,\wt{\omega}) \;=\;
\frac{t^*_\omega t_{\wt{\omega}}}{(2\pi)^2} \frac{1}{(\omega \wt{\omega})^{1/2}} \Gamma\left(1+i\frac{\omega}{\kappa}\right) \Gamma(1-i\frac{\wt{\omega}}{\kappa}) e^{-\frac{\pi}{\kappa}\frac{\omega+\wt{\omega}}{2}}   \cr &&
\;\;\;\;\;\times\;\int\limits_0^\infty d \omega'' \int\limits_0^\infty d \omega' \ \frac{ R_S}{2\pi} C_{BH}\ e^{-\frac{(\omega'-\omega'')^2}{4} R_S^2 C_{BH}}\
(\omega')^{-1/2-i\frac{\omega}{\kappa}}
(\omega'')^{-1/2+i\frac{\wt{\omega}}{\kappa}}\;.\hspace{.5in}
\label{rhosc}
\eea

Here, however, we encounter a problem: For $\;\omega=\wt{\omega}\;$, the integral in Eq.~(\ref{rhosc}) is given by
\bea
{\cal I}
&=&\int\limits_0^\infty d \omega'' \int\limits_0^\infty d \omega' \;
\frac{ R_S}{2\pi} C_{BH}\ e^{-\frac{(\omega'-\omega'')^2}{4} R_S^2 C_{BH}}\
(\omega')^{-1/2-i\frac{\omega}{\kappa}}
(\omega'')^{-1/2+i\frac{{\omega}}{\kappa}}\;,\hspace{.5in}
\label{Int1}
\eea
which diverges logarithmically  on the line $\;\omega'=\omega''\;$,
$\;
{\cal I}\propto\int\limits_0^\infty d \omega' \
(\omega')^{-1}\;.
$
But, if the diagonal of $\rho_{SC}$ diverges logarithmically, how can it be a small correction to the Hawking density matrix? The problem gets resolved by looking at the integral for  $\omega \ne \wt{\omega}$. One can then see that the diagonal is proportional to $\delta(\omega-\wt{\omega})$ plus a correction. The  $\delta(\omega-\wt{\omega})$ gets absorbed as a small contributor of order $C_{BH}^{1/2}$ into the Hawking matrix and the remaining correction  is
what constitutes the off-diagonal matrix elements. We still need to devise a subtraction scheme to isolate the diagonal part. This will be done later.

To evaluate ${\cal I}$ in Eq.~(\ref{Int1}), we will  change variables to $\;X= \omega'+\omega''\;$, $\;Y=\omega'-\omega''\;$. Since these satisfy $\;X\ge|Y|\;$, \be
\int\limits_0^\infty d \omega'' \int\limits_0^\infty d \omega' \;=\;\frac{1}{2} \int\limits_{-\infty}^\infty d Y \int\limits_{|Y|}^\infty  dX \;,
\ee
and so
\bea
&&\hspace{-1in}{\cal I}\;=\;\frac{ R_S}{2\pi} C_{BH}\frac{1}{2} \int\limits_{-\infty}^\infty d Y \ e^{-\frac{Y^2}{4} R_S^2 C_{BH}}\
\int\limits_{|Y|}^\infty  dX \left(\frac{X+Y}{2}\right)^{-1/2-i\frac{\omega}{\kappa}}
\left(\frac{X-Y}{2}\right)^{-1/2+i\frac{\wt{\omega}}{\kappa}}
\cr &=& \frac{ R_S}{2\pi} C_{BH}
\left(\frac{1}{2}\right)^{-i\frac{\omega-\wt{\omega}}{\kappa}}
\int\limits_{-\infty}^\infty d Y \ e^{-\frac{Y^2}{4} R_S^2 C_{BH}}\ Y^{-i\frac{\omega-\wt{\omega}}{\kappa}} \cr &\times&
\int\limits_{|Y|}^\infty  \frac{dX}{Y} \left(\frac{X}{Y}+1\right)^{-1/2-i\frac{\omega}{\kappa}}
\left(\frac{X}{Y}-1\right)^{-1/2+i\frac{\wt{\omega}}{\kappa}} \;.
\eea

Let us now separate the above integrations into two distinct cases, depending
on whether $Y$ is positive or negative,
\bea
&&\int\limits_{0}^\infty d Y \ e^{-\frac{Y^2}{4} R_S^2 C_{BH}}\ Y^{-i\frac{\omega-\wt{\omega}}{\kappa}}
\int\limits_{Y}^\infty  \frac{dX}{Y} \left(\frac{X}{Y}+1\right)^{-1/2-i\frac{\omega}{\kappa}}
\left(\frac{X}{Y}-1\right)^{-1/2+i\frac{\wt{\omega}}{\kappa}}
\cr &+&
\int\limits_{-\infty}^0 d Y \ e^{-\frac{Y^2}{4} R_S^2 C_{BH}}\ Y^{-i\frac{\omega-\wt{\omega}}{\kappa}}
\int\limits_{-Y}^\infty  \frac{dX}{Y} \left(\frac{X}{Y}+1\right)^{-1/2-i\frac{\omega}{\kappa}}
\left(\frac{X}{Y}-1\right)^{-1/2+i\frac{\wt{\omega}}{\kappa}} \;. \hspace{.5in}
\eea
A further change of  variables  $\;X \to Z=X/Y\;$ (and
keeping track of minus  signs) then leads to
\bea
&&\hspace{-1in} {\cal I}\;=\; \frac{ R_S}{2\pi} C_{BH}
\left(\frac{1}{2}\right)^{-i\frac{\omega-\wt{\omega}}{\kappa}}
\int\limits_{0}^\infty d Y \ e^{-\frac{Y^2}{4} R_S^2 C_{BH}}\ Y^{-i\frac{\omega-\wt{\omega}}{\kappa}} \cr && \hspace{-1in}
\;\times\;\left[
\int\limits_{1}^\infty  dZ \left(Z+1\right)^{-1/2-i\frac{\omega}{\kappa}}
\left(Z-1\right)^{-1/2+i\frac{\wt{\omega}}{\kappa}}
+\int\limits_{1}^\infty  dZ \left(Z-1\right)^{-1/2-i\frac{\omega}{\kappa}}
\left(Z+1\right)^{-1/2+i\frac{\wt{\omega}}{\kappa}}
\right]\;.\hspace{1in}
\label{Zints}
\eea
The two $Z$ integrals are related to one another under the exchange $\;\omega,\wt{\omega}\leftrightarrow -\wt{\omega},-\omega\;$, so that the full integral is invariant under the same exchange.

We now focus on just the $Z$ integrals in Eq.~(\ref{Zints}) and observe that these are
readily solved by changing variables to  $\;Z\to \wt{Z}= 1/Z\;$. The outcomes are
\be
J(\omega,\wt{\omega})\;\equiv\;\int\limits_{1}^\infty  dZ \left(Z+1\right)^{-1/2-i\frac{\omega}{\kappa}}
\left(Z-1\right)^{-1/2+i\frac{\wt{\omega}}{\kappa}}\;=\;
2^{-i \frac{\omega-\wt{\omega}}{\kappa}}\Gamma \left(i \frac{\omega-\wt{\omega}}{\kappa}\right)
\frac{\Gamma \left(\frac{1}{2}+i\frac{\wt{\omega}}{\kappa}\right)}
{\Gamma \left(\frac{1}{2}+i \frac{{\omega}}{\kappa}\right)}
\ee
for the first term and
$J(-\wt{\omega},-{\omega})$ for the second term.

It can also be observed that either of these $Z$ integrals does indeed contain
the anticipated delta function $\delta(\omega-\wt{\omega})$; for instance,
\bea
J(\omega,\wt{\omega})&=&\int\limits_{1}^\infty  dZ \left(Z-1\right)^{-1/2-i\frac{\omega}{\kappa}}
\left(Z+1\right)^{-1/2+i\frac{\wt{\omega}}{\kappa}} \cr &=&
\int\limits_{1}^\infty  dZ \frac{1}{\sqrt{Z^2-1}} \left(Z-1\right)^{-i\frac{\omega}{\kappa}}
\left(Z+1\right)^{+i\frac{\wt{\omega}}{\kappa}} \cr &=&
\int\limits_{1}^\infty  \frac{dZ}{Z}\ \left(Z\right)^{-i\frac{\omega-\wt{\omega}}{\kappa}}
+\Delta \cr &=&
\int\limits_{-\infty}^\infty  d\ln Z\ e^{-i\ln Z \frac{\omega-\wt{\omega}}{\kappa}}+\Delta \;=\; 2\pi \delta\left(\frac{\omega-\wt{\omega}}{\kappa}\right)+\Delta\;,
\eea
where  $\Delta$ denotes the finite correction to the delta function.

Evaluated as a complex integral, the  delta function is simply
\bea
\int\limits_{1}^\infty  \frac{dZ}{Z} \left(Z\right)^{-i\frac{\omega-\wt{\omega}}{\kappa}} \;=\;
-\left.\frac{z^{-i\frac{\omega-\wt{\omega}}{\kappa}}}{i \frac{\omega-\wt{\omega}}{\kappa}}\right|_1^{\infty} = \frac{- i }{\frac{\omega-\wt{\omega}}{\kappa}}\;,
\eea
so that  the regularized expression for the integral $J(\omega,\wt{\omega})$
goes as
\bea
J_{\text{reg}}(\omega,\wt{\omega})&=&\int\limits_{1}^\infty  dZ \left(Z-1\right)^{-1/2-i\frac{\omega}{\kappa}}
\left(Z+1\right)^{-1/2+i\frac{\wt{\omega}}{\kappa}}+\frac{ i }{\frac{\omega-\wt{\omega}}{\kappa}}\cr &=&
2^{-i \frac{\omega-\wt{\omega}}{\kappa}}\Gamma \left(i \frac{\omega-\wt{\omega}}{\kappa}\right)
\frac{\Gamma \left(\frac{1}{2}-i\frac{\omega}{\kappa}\right)}
{\Gamma \left(\frac{1}{2}-i \frac{\wt{\omega}}{\kappa}\right)}+\frac{ i }{\frac{\omega-\wt{\omega}}{\kappa}} \;
\eea
and similarly for
$J(-\omega,-\wt{\omega})$.

Finally, the $Y$ integral gives
\be
\int\limits_{0}^\infty d Y \ e^{-\frac{Y^2}{4} R_S^2 C_{BH}}\ Y^{-i\frac{\omega-\wt{\omega}}{\kappa}} \;=\;
\left(R_S^2 C_{BH}/4\right)^{-1/2+i \frac{\omega-\wt{\omega}}{\kappa}}
\Gamma\left(\frac{1}{2}-\frac{i}{2} \frac{\omega-\wt{\omega}}{\kappa}\right)\;.
\ee

Putting everything together, we arrive at
\bea
&&\rho_{SC}(\omega,\wt{\omega}~;C_{BH}) \;=\;
\frac{t^*_\omega t_{\wt{\omega}}}{(2\pi)^3} 2 C_{BH}^{1/2} \left(R_S^2 C_{BH}/4\right)^{+i \frac{\omega-\wt{\omega}}{\kappa}} \cr &\times&
\frac{1}{(\omega \wt{\omega})^{1/2}} \Gamma\left(1+i\frac{\omega}{\kappa}\right) \Gamma(1-i\frac{\wt{\omega}}{\kappa}) \ \hbox{\Large\em e}^{\hbox{\Large $-\frac{\pi}{\kappa}\frac{\omega+\wt{\omega}}{2}$}} \ \Gamma\left(\frac{1}{2}-\frac{i}{2} \frac{\omega-\wt{\omega}}{\kappa}\right)\cr
&\times&\Biggl\{
\Gamma \left(i \frac{\omega-\wt{\omega}}{\kappa}\right)\left[
\frac{\Gamma \left(\frac{1}{2}+i\frac{\wt{\omega}}{\kappa}\right)}
{\Gamma \left(\frac{1}{2}+i \frac{{\omega}}{\kappa}\right)}
+
\frac{\Gamma \left(\frac{1}{2}-i\frac{\omega}{\kappa}\right)}
{\Gamma \left(\frac{1}{2}-i \frac{\wt{\omega}}{\kappa}\right)}
\right] + 2\frac{ i }{\frac{\omega-\wt{\omega}}{\kappa}}\Biggr\}\;.
\label{rhoscf1}
\eea
This result has the same dimensionality
of $\omega^{-1}$ as the Hawking density matrix, given that the right-most factor on the top line has (as previously discussed) been properly regularized.

Comparing the Hawking density matrix to ours, we observe that ours contains non-trivial off-diagonal elements of order $C_{BH}^{1/2}$. Because perturbative off-diagonal elements can only enter into physical (traced) quantities like the von Neumann entropy at second order (as shown later), the implication is that corrections first appear at linear order in $C_{BH}$ as expected.

The  frequencies $\omega$ or $\wt\omega$  appear in a ratio with the temperature $\;T_H=\kappa/2\pi\hbar\;$. And so the frequencies, even when off the diagonal, are still constrained within the same ``thermal window''  as for Hawking's density matrix. In essence,  the non-vanishing elements of $\rho_{\omega\widetilde{\omega}}$  are exponentially confined to a  $T_{H}\times T_{H}$ square matrix.

We would like to point out that the quantum spread of the shell wavefunction $\sigma^2$, as defined in Eq.~(\ref{prewave}), enters Eq.~(\ref{rhoscf1})
for $\rho_{SC}$ in a very simple way: $\rho_{SC}$ is proportional to $\sigma$.
And, although  our calculations make use of  a specific proportionality constant as specified in Eq.~(\ref{wave}), the exact value of this constant will not affect our conclusions in a significant way. The effect of changing this proportionality constant will result in a different multiplicative constant of order unity in  $\rho_{SC}$.

Finally, let us reiterate that an off-diagonal density matrix is critical to any resolution of the information-loss paradox.  The off-diagonal elements are indeed the only viable place for information to be stored, as these enable the mixing of modes and allow for  non-trivial phases.

\subsection{Subleading contribution to the semiclassical density matrix}

So far, we have only  been considering contributions
to the density matrix of the form
$\beta_{\omega^{\prime}\omega}^{\ast}\beta_{\omega^{\prime\prime}\widetilde{\omega}}
e^{iv(\omega^{\prime}-\omega^{\prime\prime})}$ and neglected other possibilities; in
particular,
$\alpha_{\omega^{\prime}\omega}^{\ast}\beta_{\omega^{\prime\prime}\widetilde{\omega}}
e^{-iv(\omega^{\prime}+\omega^{\prime\prime})}$  and its complex conjugate. That such forms do not contribute to Hawking's calculation follows from  the integration over $v$  leading to
$\delta(\omega^{\prime}+\omega^{\prime\prime})$, which is necessarily vanishing because both frequencies are positive. One then might wonder if this situation can change given that, in our case, the integration over $v$ does not simply produce a delta function.

To see that this basic outcome is preserved, consider that  the only meaningful revision from our previous  integration over $v$ would be to replace the Gaussian $e^{-(\omega^{\prime}-\omega^{\prime\prime})R_S^2 C_{BH}}$ with $e^{-(\omega^{\prime}+\omega^{\prime\prime})R_S^2 C_{BH}}$. The ensuing  integration over $\;X=\omega^{\prime}+\omega^{\prime\prime}\;$ then becomes a Gaussian integral centered about $\;X=0\;$ but with integration limits from $|Y|$ to infinity. As a consequence, the integration over $X$ acts  like a delta function for
$\;Y=\omega^{\prime}-\omega^{\prime\prime}\;$. Meaning that, up to exponentially small corrections, the only contribution will be when  $\;X=Y=0\;$ or when all ingoing frequencies are vanishing. Such a case of non-propagating waves is not relevant to the process of Hawking radiation.

Hawking also comments in \cite{info} on the possibility of particle creation
due to mode mixing between rays  that do pass through the collapsing shell
(as described by Eq.~(\ref{prestuff})) and those that do not. The latter
are described by the usual plane-wave forms $\sim e^{iu\omega}$ and are
associated with trivial Bogolubov coefficients $\;\alpha_{\omega^{\prime}\omega}
\sim \delta_{\omega^{\prime},\omega}\;$,
$\;\beta_{\omega^{\prime}\omega}\sim 0\;$. As discussed in \cite{info},
such contributions can be dismissed because of the rapid variations in the
phases of the plane waves  at late retarded times, $\;u\to\infty\;$.
As $\Psi_{shell}$ has no opportunity to interfere with such modes, we can
expect Hawking's  argument to carried through unfettered.

\subsection{Multi-particle density matrix}

Until now, we have focused our attention on the density matrix for  a single particle, $\rho(\omega,\widetilde{\omega})$. In the case of $N$ identical, independent particles, it is more appropriate to work with a multi-particle density
matrix consisting of $N\times N$ blocks:  $\rho^{(N)}_{IJ}(\omega,\widetilde{\omega})$ with $\;I,J=1,\dots,N\;$.  Here,  each of
the $N^2$ entries  is itself  a matrix  having  the same dimensionality as the single-particle density matrix $\rho(\omega,\widetilde{\omega})$. In particular, each diagonal entry is the same as that of a single particle, $\;\rho^{(N)}_{II}= \rho_H(\omega,\widetilde{\omega})\;$  (where $\rho_H$ now means the
original Hawking matrix plus its subdominant semiclassical corrections
which can typically be neglected)
while each off-diagonal element contains the semiclassical part of $\rho_{\omega\widetilde{\omega}}$ only, $\;\rho^{(N)}_{I\neq J}=C^{1/2}_{BH}\Delta\rho_{OD}(\omega,\widetilde{\omega})\;$.

For the $N$-particle treatment, it is convenient to use a normalized Hawking density matrix, $\;{\rho}_H \to \frac{1}{{\rm Tr}\; \rho_H} \rho_H\;$,
where $\;{\rm Tr}\; \rho_H= \int d\omega\; \rho_H(\omega,\omega)\;$ as defined previously. It will be implied  that $\rho_H$ has  been normalized
in this way.

The $N$-particle density matrix can then be expressed as follows:
\be
\rho^{(N)}_{IJ}(\omega,\widetilde{\omega}) \;=\; \frac{1}{N}
\rho_H(\omega,\widetilde{\omega})\mathbb{I}_{N\times N}
\;+\; \frac{1}{N}
C^{1/2}_{BH}\Delta\rho_{OD}(\omega,\widetilde{\omega}) \slashed{\mathbb{I}}_{N\times N} \;.
\label{nmatrix}
\ee
The symbol $\slashed{\mathbb{I}}_{N\times N}$ denotes an $N\times N$ matrix of ones, except with the diagonal entries set to zero,
\be
\slashed{\mathbb{I}}_{N\times N}=\left[\begin{array}{ccccc}
0 & 1 & 1 &  \cdots & 1 \\
1 & 0 & 1 &  \cdots & 1 \\
1 & 1 & 0 &        & 1 \\
\vdots & \vdots & & \ddots  & \vdots \\
1 & 1 & 1 & \cdots & 0 \\
\end{array}\right]_{N\times N}\;.
\ee
The common  factor of $1/N$  ensures that the trace of the block matrix is
unity  and this normalization condition is sufficient to fix the $N$
dependence of the elements (however, see the next paragraph).

On general grounds,  each entry in the off-diagonal matrix in Eq.~(\ref{nmatrix})  is  expected to contain some  (possibly $N$-dependent) phase $e^{i\Theta_{IJ}}$. These phases result from evolution effects or from the specific initial state of the matter. There could also be additional phases that encode information about the  initial state of the collapsing shell. However, the multi-particle matrix must still be Hermitian, meaning that $\;\Theta_{IJ}=-\Theta_{JI}\;$,  and
 it can, in principle,  be diagonalized such that its real eigenvalues contain all the relevant information about the state. Of course, to diagonalize the matrix, one has to keep track of all the phases, an enormously difficult task. It follows that, from the perspective of an ``uninformed observer''({\em i.e.}, one who knows nothing about this initial state), the distribution of  phases can be regarded as random.

\section{The rate of information release}

It is clear that information is stored in the correlations that are encoded in the off-diagonal entries of the $N$-particle density matrix. Our next objective is to determine the rate in which this  information is released out of the evaporating BH.

The definition of information that we will use is the standard one: It is a
measure of how much the actual entropy $S$ deviates from the maximal, thermal  entropy of the Hawking radiation $S_H$,
\be
I=S_{H}-S.
\label{infocontent}
\ee

The thermal entropy is given by
\be \hspace{-.25in}
\frac{S_{H}}{N} \; =\; -\frac{1}{N} {\rm Tr}\left[ \mathbb{I}_{N\times N}~
\rho_H(\omega,\widetilde{\omega}) \ln\rho_H(\omega,\widetilde{\omega}) \right]=-
{\rm Tr}\left[\rho_H(\omega,\widetilde{\omega}) \ln\rho_H(\omega,\widetilde{\omega})\right]\;,
\label{S_H}
\ee
and the factor of $N^{-1}$ on the left-hand side is because this block density  matrix has been normalized so as to yield the entropy per particle. Additionally, the argument of the logarithms in Eq.~(\ref{S_H}) should  go as $\rho_H/N$ rather than just $\rho_H$. However, this logarithmic dependence on  $N$ is eliminated once the resolution of Gibbs' paradox for indistinguishable particles is taken into account.

The actual entropy is given by
\be
\;\frac{S}{N}
= -{\rm Tr}\left[\rho^{(N)}\ln{\rho^{(N)}}\right]\;,
\ee
where
$\rho^{(N)}$ is shorthand for the matrix $\rho^{(N)}_{IJ}(\omega,\widetilde{\omega})$ of Eq.~(\ref{nmatrix}).

Since $C_{BH}$ is small, the entropy can be evaluated   perturbatively,
\bea
\ \hspace{-.5in}\frac{S}{N} & = & -
\frac{1}{N}{\rm Tr}\left[\left(
\rho_H\mathbb{I}_{N\times N}
\;+\;  C^{1/2}_{BH}\Delta\rho_{OD} \slashed{\mathbb{I}}_{N\times N}\right)\ln\left(
\rho_H\mathbb{I}_{N\times N}
\;+\;  C^{1/2}_{BH}\Delta\rho_{OD} \slashed{\mathbb{I}}_{N\times N} \right)\right]
\nonumber \\
&=&
 -\frac{1}{N} {\rm Tr}\left[ \mathbb{I}_{N\times N}
\;\rho_H \ln\rho_H \right]\; -\;  \frac{1}{N} {C_{BH}} {\rm Tr}\left[ \left(\Delta\rho_{OD} \slashed{\mathbb{I}}_{N\times N}\right)^2 (\rho_H\ \mathbb{I}_{N\times N})^{-1} \right] \nonumber \\
&\;&\;\;\;\;\;+\; \frac{C_{BH}}{2} \frac{1}{N} {\rm Tr}\left[(\rho_H\ \mathbb{I}_{N\times N}) \left(\Delta\rho_{OD} \slashed{\mathbb{I}}_{N\times N}\right)^2 (\rho_H\ \mathbb{I}_{N\times N})^{-2}   \right]
\;+\; \cdots \cr &=& -
{\rm Tr}\left[\rho_H \ln\rho_H\right] \;-\; \frac{1}{2} N C_{BH} {\rm Tr}\left[ (\Delta\rho_{OD})^2 \rho_H^{-1} \right]
\;+\;\cdots\;,
\eea
with the leading-order correction going linear in $C_{BH}$ as already claimed.
Here, we have used that $\;{\rm Tr}\left[\slashed{\mathbb{I}}_{N\times N}\right]=0\;$ and $\;{\rm Tr}\left[
\slashed{\mathbb{I}}_{N\times N}^2\right]=N(N-1)\sim N^2\;$.

Therefore, up to corrections,
\bea
S&=&S_H\left(1- \frac{1}{2} N C_{BH} \frac{{\rm Tr}\left[ (\Delta\rho_{OD})^2 \rho_H^{-1} \right]}{-{\rm Tr}\left[
\rho_H \ln\rho_H\  \right]}\right) \cr &=&
S_H\left(1- \frac{1}{2} K\ N C_{BH}\right)\;,
\label{staylor}
\eea
where $K=\frac{{\rm Tr}\left[ (\Delta\rho_{OD})^2 \rho_H^{-1} \right]}{-{\rm Tr}\left[
\rho_H \ln\rho_H\ \right]}$ is a positive numerical factor of order one.

We observe from Eq.~(\ref{staylor}) that the expansion parameter is $N C_{BH}$. As long as $\;N C_{BH} <1\;$, we can formally expand the expression for $S$,
treating the off-diagonal part as a perturbation of
the dominant Hawking contribution. However, once $\;N C_{BH} =1\;$, the expansion
breaks down, and  this special point happens to coincide with the Page time.

To substantiate this last assertion, let us discuss the time dependence of $N$ and $C_{BH}$. We do so only approximately by substituting
their classical expressions
 into Eqs.~(\ref{infocontent}) and (\ref{staylor}). This corresponds to substituting $\langle N\rangle$ and $\langle C_{BH}\rangle$ rather than evaluating the actual time-dependent density matrix. Nevertheless, for a rough, qualitative estimate of the time-dependence of $I$, this will suffice.

Let us first estimate $N(t)$, the total  number of emitted particles up to any  given time.
Now, the number of emitted photons (or other massless particles)
can be deduced
by starting with  $\;dN\simeq  dM\frac{dN}{dM}=-\frac{dM}{T_{H}}\;$ or,
in terms of $R_{S}$, $\;dN\simeq -\frac{2\pi}{G\hbar}R_SdR_S\;$ and so
$\;N(t)\simeq -\frac{2\pi}{G\hbar}\int\limits^{R_S(t)}_{R_S(0)} dR_{S}\; R_{S}\;$.
Integrating and applying the
definition of $S_{BH}$, we then obtain
\be
N(t)\;\simeq\;  S_{BH}(0) - S_{BH}(t)\;.
\ee
But the right hand-side is simply the (thermal) entropy lost to radiation,
so that
\be
N(t)\;\simeq\; S_H(t) \;.
\ee

The Page time $t_{Page}$ is formally defined
as the moment  when the entropy lost to radiation is
one half of the initial BH entropy \cite{page}, meaning that
$\;N(t_{Page})\simeq \frac{1}{2}S_{BH}(0)\;$. At the same time,
$\;C_{BH}(t_{Page})=\frac{1}{S_{BH}(t_{Page})}=\frac{2}{S_{BH}(0)}\;$.
We thus have $\;N(t_{Page})C_{BH}(t_{Page})\simeq 1 \;$ as claimed.

Let us now  reconsider the information content of the radiation by way
of
Eqs.~(\ref{infocontent}) and~(\ref{staylor}).
As $\;N(t)\simeq S_H(t)\;$ and $\;C_{BH}(t)=S_{BH}^{-1}(t)\simeq\left(S_{BH}(0)-S_H(t)\right)^{-1}\;$,  this is
\bea
I(t)\;=\; S_H(t)-S(t) \;&=&\; \frac{K}{2}S_H(t) N(t) C_{BH}(t) \cr &\simeq& \frac{K}{2}\frac{S_H(t)^2}{S_{BH}(0)-S_H(t)}
\;.
\eea
It follows that
\bea
\frac{dI}{dS_H}&\simeq &
\frac{K}{2}\left[\frac{2S_H}{S_{BH}(0)-S_H}+\frac{S_H^2}{\left(S_{BH}(0)-S_H\right)^2}\right]\nonumber \\
&\simeq&\frac{K}{2}\left[2+C_{BH}N\right]
C_{BH}N\;.
\label{inforate}
\eea

We are thus lead to the following picture. At early times in the evaporation process ({\em i.e.}, before the Page time, $\;NC_{BH}<1\;$), the released information is
perturbatively small,
\be
\left.\frac{dI}{dS_{H}}\right|_{t<t_{Page}}\;\sim\; {\cal O}[C_{BH}]\;.
\ee
However, at the Page time and beyond, the right-hand of Eq.~(\ref{inforate})
is of order unity,
\be
\left.\frac{dI}{dS_{H}}\right|_{t>t_{Page}}\; \sim \; 1 \;.
\ee

This picture is qualitatively similar to the dependence that Page has found  in the second paper of \cite{page}. However, the initial rate of information release that we find is higher than Page's. For $\;t>t_{Page}\;$, the expansion in $N C_{BH}$ breaks down and our result is of limited value. We do, however, expect that the late rate of information release is saturated by unity as Page has found.

For a measure of the purity of the state of the radiation, one can look at the ratio $\frac{{\rm Tr}\left[\left(\rho^{(N)}\right)^2\right]}{\left({\rm Tr}\; \rho^{(N)}\right)^2}$.
But, because of the normalization of the density matrix $\;{\rm Tr}\; \rho^{(N)} =1\;$, we need only consider  the magnitude of the numerator,
\be
{\rm Tr} \left[\left(\rho^{(N)}\right)^2\right] \;\simeq\;
\frac{1}{N^2}{\rm Tr}\left[ N \rho_H^2+ N^2 C_{BH} (\Delta\rho_{OD})^2 \right]\;,
\ee
where $\;{\rm Tr}\;\slashed{\mathbb{I}}_{N\times N}=0\;$
and $\;{\rm Tr}\;\slashed{\mathbb{I}}^2_{N\times N} = N(N-1)\sim N^2\;$
have again been applied.
Consequently,
\be
\frac{{\rm Tr} \left[\left(\rho^{(N)}\right)^2\right]}{\left({\rm Tr}\; \rho^{(N)}\right)^2} \;\simeq\;
\frac{1}{N}  {\rm Tr}\;\rho_H^2\left(1+ N C_{BH} \frac{ {\rm Tr}\left[\Delta\rho_{OD}\right]^2 }{ {\rm Tr}\;\rho_H^2} \right)\;.
\label{ratio}
\ee

One can  see that, at the Page time when $\;N C_{BH}=1\;$, the deviation of the ratio from its Hawking value becomes significant. The density matrix is still very close to thermal, but it is already possible to distinguish the actual matrix from a maximally mixed  one with order $1/N$ accuracy rather than exponential accuracy.

It can also be seen  that, as the evaporation proceeds, the trend is consistently away from thermality.
We can define the ``rate of purification'' as the rate in which the deviation from a thermal matrix increases. It is evident that this rate is proportional to the rate of change of $N C_{BH}$. As discussed above,  $N$ and $C_{BH}$ both increase monotonically with time. Equivalently, these  parameters both increase linearly with energy emitted from the BH as follows from the first law of thermodynamics; meaning  that the rate of purification is increasing quadratically so. This  is  consistent with the expectation that the full information content will be released eventually.

Let us emphasize that the full release of information could only be deduced  by an observer who is privy to the initial state of the matter or has managed to collect (and then analyze) an order-one fraction of all the particles that will be radiated by the BH. Nonetheless, this state of affairs is no different than it would be for someone trying to reassemble the original information contained within a burning encyclopedia.

\section{Conclusion}

We have shown how quantum fluctuations in the background geometry that are induced by the quantum fluctuations of the collapsing shell lead to
a density matrix for BH radiation that is different from
that of Hawking's. The fundamental difference is
the presence of small off-diagonal elements, which
provides the necessary receptacle for information storage. The magnitude of the off-diagonal elements is determined by the strength of the quantum fluctuations of the shell as it approaches its Schwarzschild radius. At  late times, when a finite fraction of the BH has evaporated,  the off-diagonal components gain dominance and allow the state of the radiation to purify at an increasing rate. 

We interpret our results as an indication that quantum gravity does not introduce an additional layer of unpredictability, as claimed by Hawking. When the background geometry is treated according to the standard rules of quantum mechanics, the results are consistent with the normal  expectations.

A shortcoming of our analysis is as follows: Whereas the
Hawking particles have been
emitted at different times throughout the evaporation process,
our formalism does not directly account for this fact. Hence,
it could well be  that the time dependence of
$N(t)$ and $C_{BH}(t)$ is  more complicated than
insinuated by the previous analysis. A more accurate treatment
would require the inclusion of time dependence in
(at least) the  BH wavefunction and basis
functions $e^{iv\omega'}$. We intend to address this
matter at a later time \cite{last}.

\section*{Acknowledgments}

We thank Merav Hadad for discussions and Sunny Itzhaki and Amos Yarom for useful comments on the manuscript. The research of RB was supported by the Israel Science Foundation grant no. 239/10. The research of AJMM received support from a Rhodes University Discretionary Grant RD35/2012.
AJMM thanks Ben Gurion University for their hospitality during the initial stages of this project.

\appendix

\end{document}